# Grain boundaries in ultrafine grained materials processed by severe plastic deformation and related phenomena

X. Sauvage[1], G. Wilde[2], S.V. Divinski[2], Z. Horita[3], R.Z. Valiev[4]

*1-University of Rouen, CNRS UMR 6634, Groupe de Physique des Matériaux, Faculté des Sciences, BP12, 76801 Saint-Etienne du Rouvray, France*

*2- Institute of Materials Physics, University of Münster, Münster, Germany*

*3- Department of Materials Science and Engineering, Faculty of Engineering, Kyushu University, Fukuoka 819-0395, Japan*

*4- Institute for Physics of Advanced Materials, Ufa State Aviation Technical University, K. Marx 12, Ufa 450000, Russia*

**Abstract**

Grain boundaries in ultrafine grained (UFG) materials processed by severe plastic deformation (SPD) are often called "non-equilibrium" grain boundaries. Such boundaries are characterized by excess grain boundary energy, presence of long range elastic stresses and enhanced free volumes. These features and related phenomena (diffusion, segregation, etc.) have been the object of intense studies and the obtained results provide convincing evidence of the importance of a non-equilibrium state of high angle grain boundaries for UFG materials with unusual properties. The aims of the present paper are first to give a short overview of this research field and then to consider tangled, yet unclear issues and outline the ways of oncoming studies. A special emphasis is given on the specific structure of grain boundaries in ultrafine grained materials processed by SPD, on grain boundary segregation, on interfacial mixing linked to heterophase boundaries and on grain boundary diffusion. The connection between these unique features and the mechanical properties or the thermal stability of the ultrafine grained alloys is also discussed.

# 1. Introduction

With grain sizes in a submicron (100 – 1000 nm) or nanocrystalline (< 100 nm) range ultrafine-grained (UFG) materials contain in their microstructure a very high density of grain boundaries (GB), which can play a significant role in the development and exhibition of novel properties. For this reason, UFG materials can be typically considered as interface-controlled materials [1]. Unlikely to the nanocrystalline materials where grain boundary material can represent a significant, e.g. a per cent or even larger fraction of the whole volume, the volume fraction of GBs in an UFG material is less than 1%. However, the structure, kinetic and thermodynamic properties of GBs could be modified so significantly that their start to dominate some important material properties.

Already in first works on nanocrystalline materials pioneered by Gleiter and colleagues it was suggested that grain boundaries can possess a number of peculiar features in terms of their atomic structure in contrast to grain boundaries in conventional polycrystalline materials [1,2]. Further studies delivered plenty of indications towards this idea, evidencing simultaneously the fact that solely the grain size is not the deciding parameter. For example, specific grain boundaries were revealed in ultrafine-grained materials produced by severe plastic deformation (SPD) techniques [3]. In the recent decade the use of SPD techniques for grain refinement and nanostructuring of metals and alloys attracted intensive attention and received much development due to their possibility not only to enhance properties of different materials but also to produce mulifunctionality of the materials including commercial alloys and composites and presently, these developments witness the stage of transition from laboratory research to their practical application [4-6].

Depending on the regimes of SPD processing different types of grain boundaries can be formed in the UFG materials (high- and low-angle, special and random, equilibrium and so-called "non-equilibrium" grain boundaries) [3,7], which paves the way to grain boundary engineering of UFG materials, i.e. to the control of their properties by means of varying the grain boundary structure. For example, recent studies demonstrated that transport properties of UFG materials (diffusion, segregation, etc.) are markedly affected by a so-called "non-equilibrium" grain boundary state [8-10]. At this place it is important to highlight that a broad spectrum of diffusivities of short-circuit paths is observed in UFG materials – contributions of high-angle grain boundaries with both "normal" and significantly enhanced diffusion rates can be differentiated in SPD-processed materials [11, 12]. In this context, the "normal" diffusion rates are those which reveal the relaxed general high-angle grain boundaries as they are present in well-annealed polycrystalline counterparts[1] and the non-equilibrium interfaces are characterized by considerably higher diffusion coefficients. This hierarchy of interfaces in terms of their corresponding diffusivities is proposed [12] to explain the apparent contradictions between earlier publications that reported either conventional or unusual properties for grain boundaries in nanocrystalline or ultrafine grained materials.

The notions on non-equilibrium grain boundaries were first introduced in the scientific literature in the 1980s [13, 14] reasoning from investigations of interactions of lattice dislocations with grain boundaries. According to [14] the formation of a non-equilibrium grain boundary state is characterized by three main features, namely, excess grain boundary energy (at the specified crystallographic parameters of the boundary), the presence of long range elastic stresses (Figure 1) and enhanced free volume. Discontinuous distortions of crystallographically ordered structures, that may come about by accommodation problems of differently oriented crystallites of finite sizes or by high densities of lattice dislocations and their interaction with grain boundaries can be considered as sources of elastic stress fields that modify the atomic structure of high angle grain boundaries so that their excess free energy becomes enhanced. Somewhat unfortunately, these "unusual" grain boundaries have been termed "non-equilibrium" grain boundaries although in a strict sense, each grain boundary is a non-equilibrium defect if segregation effects (see section 3) are not to be considered. Since however the term has been accepted and utilized by the entire community who works on severe plastic deformation, we will also use it here.

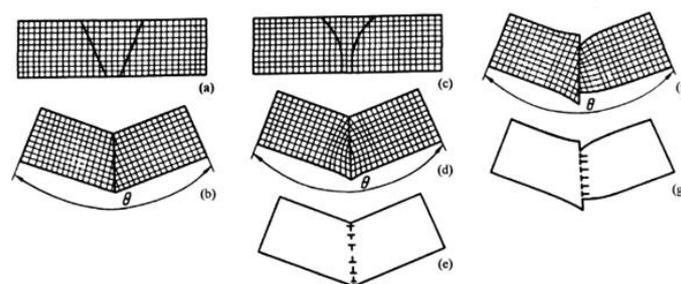

Figure 1
Phenomenological engineering of grain boundaries via thought cuts:
(a) → (b) - equilibrium grain boundary (bunches of deformation are connected without mesoscopic strain);
(c) → (d), and (a) → (f) - non-equilibrium grain boundaries (deformation is required for crystal joint –
bending and tension - compression, respectively);
(e), (g) – schemes of grain boundary dislocations (GBD) complexes initiating the same character of mesoscopic elastic distortions as in (d), (f) [14].

---

[1] Where they are typically the fastest short-circuit diffusion paths

A model for these non-equilibrium grain boundaries has been developed by A.A. Nazarov, A.E. Romanov and R.Z. Valiev in a series of papers [15, 16] describing their formation. Lattice dislocations that are created during the plastic straining move towards high angle grain boundaries on their respective glide planes during continued straining and then, when reaching a high-angle grain boundary, transform into so-called "extrinsic grain boundary dislocations", i.e. dislocations that do not contribute towards the misorientation of the two adjacent grains. As a net effect, high angle grain boundaries with high densities of such extrinsic grain boundary dislocations would also contain increased energy and free volume and considerable microstrain associated with the grain boundary region [15].

In recent years the non-equilibrium grain boundaries in UFG materials and related phenomena (diffusion, segregation, etc.) have been the object of intense studies performed by the authors of this paper and the obtained results provide convincing evidence of the importance of a non-equilibrium state of high angle grain boundaries for UFG materials with unusual properties. At the same time the complexity of such research becomes evident, involving the most contemporary techniques of structural analysis and, occasionally, different interpretation of the obtained results. All this specifies the aims of the present paper - first, to introduce the readers to this research field of recent studies of grain boundaries in bulk nanostructured materials where unique features about their structures and properties are outlined; second, to consider tangled, yet unclear issues and outline the ways of oncoming studies. The available models of the "non-equilibrium" GBs will be examined against the newest experimental data.

## 2. Structure of grain boundaries in ultrafine grained materials

The atomistic structure of random high-angle grain boundaries has been discussed since several decades by different models assuming quite different structural arrangements ranging from an amorphous structure to local structural units with high packing densities that are arranged non-periodically along the boundary plane, see e.g. [17-19], to mention just a few examples. In recent years, atomistic simulations have considerably contributed to the understanding of grain boundary structures [20-23], yet without yielding a unique description of the atomic structure of random high angle grain boundaries.

However, the goal set for the present review is not to unravel the real space arrangement of atoms within the boundary plane of random high angle grain boundaries, but analyze the structural modifications of high angle grain boundaries inflicted by severe plastic deformation processing, for formation of which during SPD processing exist already strong indications, see e.g. [3, 24]. In earlier studies of grain boundaries in UFG materials processed by SPD techniques there have been already used various, often mutually complementary, structural methods: transmission electron microscopy (TEM), X-ray diffraction, Mössbauer spectroscopy, dilatometry, differential calorimetry and others (see, e.g. [3]). They clearly evidenced that mostly high-angle grain boundaries leading to grain refinement can be formed after optimization of SPD processing routes and these grain boundaries possess specific non-equilibrium structures. Later, structure sensitive probes have been applied that are sensitive to modifications of the atomic structure, such as grain boundary diffusion measurements (see section 5) or high resolution transmission electron microscopy (HRTEM) analyses, in order to identify and characterize transformations of the grain boundary structure due to the severe deformation processing.

Z. Horita et al. [24] as well as R.Z. Valiev et al. [3] noticed serrated contrast features in bright field transmission electron microscopy (TEM) images and also in HRTEM analyses, respectively that were interpreted as evidence for a high local density of dislocation structures associated with the apparent non-equilibrium grain boundary. However, due to the possible Moiré effect occurring in the projection of sample regions near interfaces and due to the delocalization of information in HRTEM analyses by aberrations of the electromagnetic lenses, unambiguous interpretations of grain boundary structures require more sophisticated analyses, which today can be provided by the use of so-called $C_s$-corrected TEMs that are corrected for spherical aberration. Figure 2(a) shows the HRTEM image of a grain boundary with both adjacent grains oriented with their <110> zone axis parallel to the electron beam in a $Pd_{90}Ag_{10}$ alloy that had been severely deformed by repeated rolling and folding. The details of the synthesis process are similar to the procedure described earlier [25, 26]. The image was taken with a FEI-Titan TEM equipped with a field emission cathode and a $C_s$-corrector. The non-equilibrium character of this grain boundary in term of the interpretations given earlier by Z. Horita et al. or R.Z. Valiev et al. is manifested in the non-uniform faceted form (Fig. 2(a)) of the two joining <110>-oriented grains. Grain boundaries with similar features are commonly observed for materials after SPD processing [27]. Yet, it should be noted that not all grain boundaries in severely deformed materials present morphologies as in Figure 2(a). In fact, only a minority of grain boundaries with an average spacing of a few grain diameters display non-uniform faceting, implying that also during SPD processing, the localization of deformation controls the evolution of the microstructure.

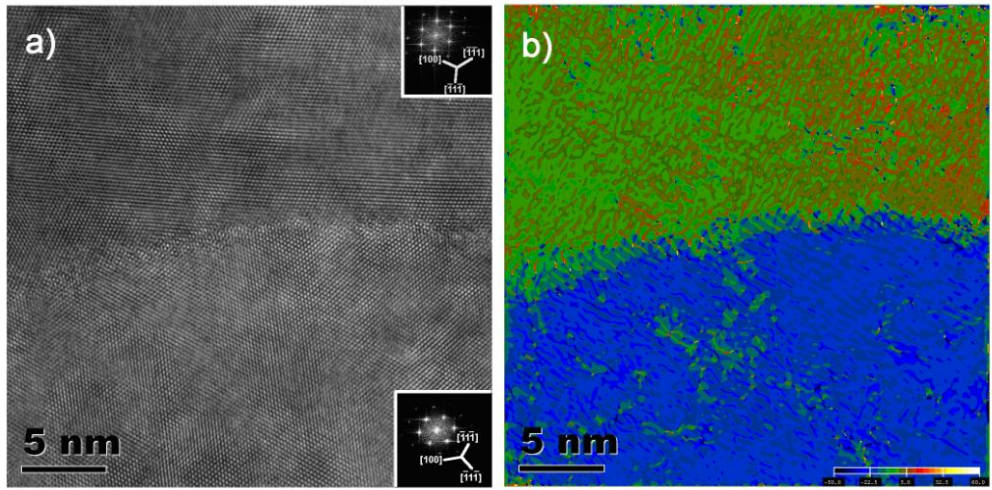

Figure 2
(a) HRTEM micrograph showing two adjacent grains of Ag10Pd90 (after 110 cycles) both oriented along the <110> direction [26]. The grains are rotated by an angle of 23.7° with respect to each other. The grain boundary appears here in a wavy faceted form. The Fourier transforms of the images of both grains are given as insets. (b) Strain map showing the in-plane rigid-body rotation xy (rotations on a scale from -50° to +60° (anticlockwise positive) are displayed. Hot spots refer to dislocation cores. The greenish colour of the upper grain represents the zero distortion taken as a reference).

In order to analyze whether the non-uniformly faceted grain boundaries might correspond to non-equilibrium grain boundaries, the residual microstrain present at the grain boundary shown in Fig. 2(a) was analyzed by the method of geometric phase analysis (GPA) that allows calculating relative magnitudes of the in-plane components of the strain tensor and of the tensor of rigid body rotation with respect to a reference lattice, based on the intensity distribution in high resolution electron micrographs [28, 29]. Details concerning the analysis and concerning the conditions under which the HRTEM micrograph was taken can be found in [9]. One result of this analysis is displayed in Fig. 2(b) as the rigid body rotation. In addition to the misorientation between the two neighbouring grains, a clear and significant variation of the colour representing the variation of the relative rotations of the lattice is observed in the near-boundary region. It should be noted, that the bright spots ("hot spots") in Fig. 2(b) represent regions where a discontinuity in the transmitted phase of the electron wave occurs, i.e. these spots mark the positions of the core region of full or partial dislocations. Additionally, local structures within the distorted grain boundary region that show an abrupt change of the local orientation of the crystal lattice with respect to the orientation of the lattice of the parent grain could also contribute such bright features in the strain maps, since the linear density of the hot spots that is estimated to be about $10^9$ m$^{-1}$ is much too high to be associated with the extrinsic GB dislocations only. By integrating the strain over rectangular regions with the long axis perpendicular to the estimated boundary plane, the width of the boundary in terms of the strain distribution was found to be in a range of 1.5 nm to 2 nm, which is about double the value for relaxed grain boundaries analyzed by the same method (The GB width was directly determined by following the rotational component of the averaged strain field around the grain boundary). The observed topology of the strain distribution at this grain boundary in the severely strained Pd-Ag alloy clearly serves to contribute an enhanced excess free energy density to the grain boundary energy, supporting the existence of non-equilibrium grain boundaries after SPD processing. This first result of strain mapping at such a grain boundary is also direct support for the interpretations of GB segregations and the diffusion studies discussed in the following chapters.

Thus, with respect to the grain boundary structure in SPD-processed materials with ultrafine grain size, recent studies enable to conclude that:

- Non-equilibrium grain boundaries exist in UFG materials and these specific grain boundaries possess an increased free energy density, increased width, high density of dislocations (full or partial) associated with the near-boundary region and correspondingly large residual microstrain.

- The structural width of non-equilibrium grain boundaries is significantly smaller than 10 nm, if the rotational component of the strain gradient across the interface is used as a measure of the width. It reaches a value of 1.5 to 2.0 nm being still twice as large as the width of relaxed high-angle grain boundaries in annealed coarse-grained materials. The shear components of the strain field reveal similar values of the grain boundary width.

The density of 'hot spots' in GPA of a non-equilibrium grain boundary is remarkably large, about $10^9$ m$^{-1}$ (Fig. 2(b)) and, thus, these features cannot be directly interpreted in terms of e.g. extrinsic grain boundary dislocations. In fact, these features in the strain map indicate the presence of distinct structural units with an abrupt change of the local orientation of the crystal lattice (in addition to full or partial dislocations), which might result from severe dislocation accumulation and/or dislocation dissociation at the grain boundary. Note that the density of "extrinsic grain boundary dislocations" was estimated based on diffusion measurements (see section 5) to be about $5 \times 10^7$ m$^{-1}$ in Ni processed by Equal Channel Angular Pressing (ECAP) [27]. Yet, these structural features need to be investigated, e.g. by comparing strain maps of grain boundaries obtained by atomistic simulations with respective results based on high resolution transmission electron microscopy.

## 3. SPD induced GB segregation

The grain size refinement mechanism during SPD is controlled by the generation of dislocations, the way they do dynamically reorganize to form low angle - and finally, for larger strains, high angle boundaries [3, 30]. On the other hand, it is also well known that impurities or solute elements may have strong interactions with dislocations. They usually lead to a stronger strain hardening due to a higher dislocation production rate during deformation [31], but alloying elements may also modify the stacking fault energy [32, 33] making twinning more or less energetically favourable. These features may explain why a small change in the alloying element concentration or in the impurity level can dramatically change the grain size achievable by SPD. This is particularly impressive in aluminium alloyed with few percent of Mg. Indeed, pure Al processed by High Pressure Torsion (HPT) leads to a grain size of 800 nm [34], while it is decreased to 150 nm in Al-3%Mg [35] and reaches even 100 nm in Al-6%Mg [36]. It is also important to note that extremely low levels of impurities may also dramatically influence the grain size achievable by SPD as reported in Al [37] and in Ni [38] but it should be noted that the effect of impurities seems less pronounced in Cu [37]. This specificity could be explained by the relatively smaller stacking fault energy of Cu, and thus a lower mobility of dislocations leading to more accumulation of dislocations to contribute to grain refinement at room temperature.

Solute elements are known to interact with all kinds of structural defects like vacancies [39], dislocations [40, 41], stacking faults [42] and grain boundaries [43]. This latter phenomenon has been widely investigated in numerous alloys because of its dramatic influence on the mechanical behaviour (creep, toughness or ductility). For example, it has been reported that in Al alloys Mg may segregate along grain boundaries [44-46], similar features were observed for Si, P and C in bcc Fe [47], while B exhibits the same behaviour in $Ni_3Al$ [48], FeAl [49], but was also detected along γ/γ' heterophase interfaces in superalloys [50]. Moreover, small quantities of residual impurities (especially of strongly segregating ones like P, S or C in Cu or Ni) can significantly modify GB diffusivities even at the ppm level [51, 52].

Indirect evidence of grain boundary segregation in UFG materials processed by SPD has been reported in a few cases where the thermal stability has been investigated as a function of the impurity level in nickel [38] or as a function of the concentration of Sb in copper [53]. However, there are only a limited number of reports providing direct evidence of grain boundary segregation in UFG materials processed by SPD. Most of them rely on atomic scale characterization thanks to Atom Probe Tomography (APT). Unfortunately, this technique provides only very limited crystallographic information and the GB misorientation is usually unknown. Moreover only small GB areas can be analyzed making any statistics almost impossible. Anyway, it was demonstrated that GB segregation in SPD materials is not a marginal feature but could be observed in various kinds of alloys.

Segregated elements could be some impurities resulting from the casting process, like O and C in titanium [56]. In the case of accumulated processes, like Accumulated Roll Bonding, even if surfaces are carefully cleaned between each step, some contamination may occur and additional impurities might be incorporated leading at the end to some significant segregations [57]. In steels processed by SPD, the progressive decomposition of carbides leads to carbon supersaturated solutions. Released carbon atoms are trapped by dislocations and significantly hinder dynamic recovery processes leading to a grain size of only 10-20 nm for pearlitic steels processed by HPT [58]. This is one order of magnitude smaller than in commercially pure Fe, demonstrating the strong influence of SPD induced segregations on the grain size refinement mechanisms. However, this specific case cannot primarily be considered as GB segregation. It is the result of solute element (for instance carbon) trapping by dislocations in the course of SPD.

Combining strengthening by grain size reduction with solid solution hardening or precipitate hardening are two attractive ways for the improvement of the mechanical properties of UFG materials. It was even proposed by some authors that a fine distribution of nanoscaled precipitates may act as sites for trapping and accumulating dislocations, leading to an increase of the strain hardening and subsequently to an increased ductility [59, 60]. The main challenge is to control the precipitation kinetics in a situation where recrystallisation, grain growth, and heterogeneous precipitation along dislocations and/or grain boundaries are very likely to occur. This approach was quite successfully pursued by Kim and co-authors in a 6061AA alloy processed by Accumulative Roll Bonding (ARB) [61, 62]. They have shown that a significant strengthening could be obtained thanks to the combination of an ultrafine grained structure and nanoscaled precipitates. Similar features together with an improved ductility were also reported by Zhao and co-authors on a 7075 AA and by Ohashi and co-authors on an Al-11wt%Ag alloy processed by ECAP followed by a precipitation treatment [63-66]. In any case, the deformed material has to contain a significant concentration of alloying elements in solid solution that could be candidate for some GB segregations. Such feature was reported in 6061 AA processed by HPT [10] or ECAP [67], where Mg, Cu and Si segregations along planar defects attributed to grain boundaries were observed. The solute element enriched layer is only about 2 nm and the local enrichment does not exceed 2 at.%. Mg segregation was also reported for an AlMgCuZn alloy processed by ECAP [68], a 7075AA processed by HPT [69] and an Al-6.8%Mg processed by HPT [70]. In this latter case, after large deformation by HPT (20 turns), a mean grain size of about 100 nm is achieved, with a large fraction of High Angle Grain Boundaries (HAGB). Using APT, very strong local enrichments up to 25 at. % in a much thicker layer (6 to 8 nm) were observed (Fig. 3). This thickness is much larger than the GB width measured on HRTEM images thanks to GPA (about 2 nm, see section 2), where this parameter was defined as a zone where the averaged rotational component of the strain field is changed from the value in one grain to that in its neighbour. It seems, a non-equilibrium GB may incorporate a larger amount of segregating atoms with respect to relaxed interfaces (as a result of the increased free volume) but the thickness of the segregated layer might not only be determined by the distorted layer near the non-equilibrium GB (as it is defined by GPA).

APT data also revealed that grain boundaries are not homogeneously covered and that the Mg concentration in solid solution may strongly vary from one grain to another [70]. Such features might indicate that the local configuration of the GB and especially dislocations lying in the vicinity of the boundaries may affect the distribution of solute elements. It is also interesting to note that this material exhibits a very high yield stress after HPT processing, much higher than a prediction based on the Hall-Petch law. Thus, it seems that GB segregations could significantly affect the deformation mechanisms (dislocation nucleation and glide) in UFG materials processed by SPD.

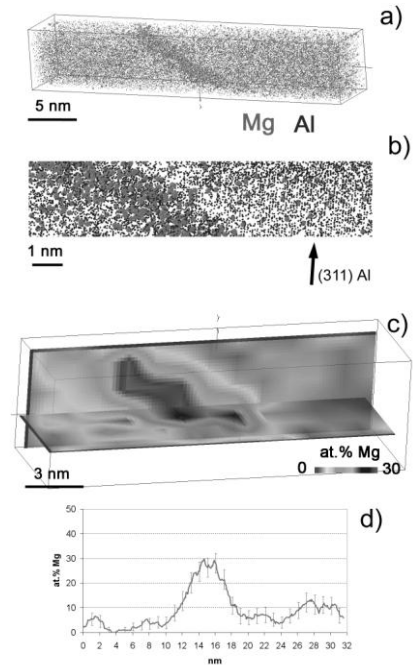

Figure 3
3-D reconstruction of an analyzed volume in the UFG 1570 alloy:
(a) full data set showing a planar segregation of Mg (Al atoms are displayed as dots and Mg atoms as bubbles); (b) selected part orientated to display (311)Al atomic planes on the right of the planar segregation; (c) 2-D chemical map showing the Mg concentration fluctuations within the volume; (d) concentration profile computed across the segregation (sampling volume thickness 1 nm). – from [70].)

The physical mechanisms of SPD induced segregation are still under debate. The driving force for GB segregations is usually the minimization of the grain boundary energy [43], however in some specific cases, some so-called non-equilibrium GB segregations may occur. If the density of vacancies is above the equilibrium value, they might diffuse towards sinks like GBs. Then, in case of a positive binding energy between a solute element and vacancies, the flux of vacancies may lead to non-equilibrium GB segregations. A high density of point defects usually arises from irradiation or deformation [71]. The latter has been recently demonstrated in the in-situ annealing experiments on HPT Cu and Ni samples [72].

The large density of point defects created during SPD may enhance the atomic mobility. In case of equilibrium grain boundary segregation, these point defects simply promote the diffusion of solute elements that would be impossible at low temperature. Dislocations may also play a significant role in enhancing the atomic mobility, through the well known pipe diffusion mechanism or solute drag if the velocity of the dislocations is not too high. Comparable to materials under irradiation, it is also reasonable to assume that a vacancy flux towards GBs acting as sinks may promote the formation of non equilibrium grain boundary segregations.

However, some irradiation experiments performed on stainless steel nanostructured by HPT have clearly demonstrated that GB segregation which did not appear during SPD could be triggered by irradiation [73]. Therefore, while there is little doubt about the large density of point defects created during SPD, the flux toward grain boundaries is probably much lower because of other sinks like dislocations activated during the process.

## 4. Heterophase boundaries and multiphase alloys during SPD

It is known since a very long time that heterophase boundaries may promote the grain size reduction during deformation and thus the resulting strengthening. This is the typical case of drawn pearlitic steels, for which an interlamellar spacing of only 20 nm is commonly achieved in mass production leading to a yield stress of up to 3 GPa or more [74-76]. Following this approach, multiphase materials containing different phases with the capability of co-deformation are of particular interest for the SPD community. There are several ways of producing heterophase boundaries. Normal casting produces multiphase structures and fine structures may be formed during solidification for the compositions corresponding to an eutectic or eutectoid reaction. However, SPD is a good tool to further break down the sizes of the dispersoids regardless of the alloying compositions [77]. The fragmentation can be achieved by intense slip of dislocations and repetitive processing of SPD refines the particle size to the nanometer levels. Examples are an Al-5wt%Cu alloy [78], Al-11wt%%Ag alloy [66] and Al-Mg-Si alloy [77]. It was shown that the grain size was reduced by the presence of such heterophase particles in the matrix [79-83] and the typical range of the grain size is 100 to 500 nm [3]. Some other multiphase materials may exhibit a grain size as small as 10 or 20 nm [84-85]. Figure 4 shows the nanoscaled structure of such a material (a CuCr composite containing 57% Cu and 43% Cr) where the grain size achieved thanks to HPT is about 20 nm. In an Fe-based alloy where the elastic modulus was designed to be low by addition of alloying elements, a phase transformation occurred during HPT and it was shown that the grain size was reduced to 20-50 nm with high strength and good ductility [86].

An artificial way of preparing a multiphase nanostructured alloy can be derived from the ARB technique, using as starting material a sandwich structure made of different metals [87, 88]. Heterophases may also be produced by mixing with different kinds of powders and consolidation by HPT or ECAP. Examples are inclusions of ceramic particles or carbon nanotubes in the metallic matrix [89-92]. Allotropic phase transformations through processing by HPT under high pressure produces heterophase boundaries even though the sample is composed of single pure metals such as Ti and Zr [93-99].

During SPD, large densities of lattice defects like dislocations or boundaries are created; pushing systems far away from the thermodynamic equilibrium state and these features may affect their stability [3, 99]. Thus, phase transformations may occur, leading to the decomposition of second phase particles and the vanishing of heterophase interfaces. It also introduces supersaturation of alloying elements and thus brings about significant hardening by solid solution [100-102]. Such features were reported in steels where $Fe_3C$ carbides were dissolved after HPT [58, 103, 104] but also in FeNi alloys [105, 106] processed by HPT or in an Al-Cu alloy processed by ECAP [78, 107]. In any case, dislocations are thought to play a major role by shearing precipitates and by dragging solute atoms.

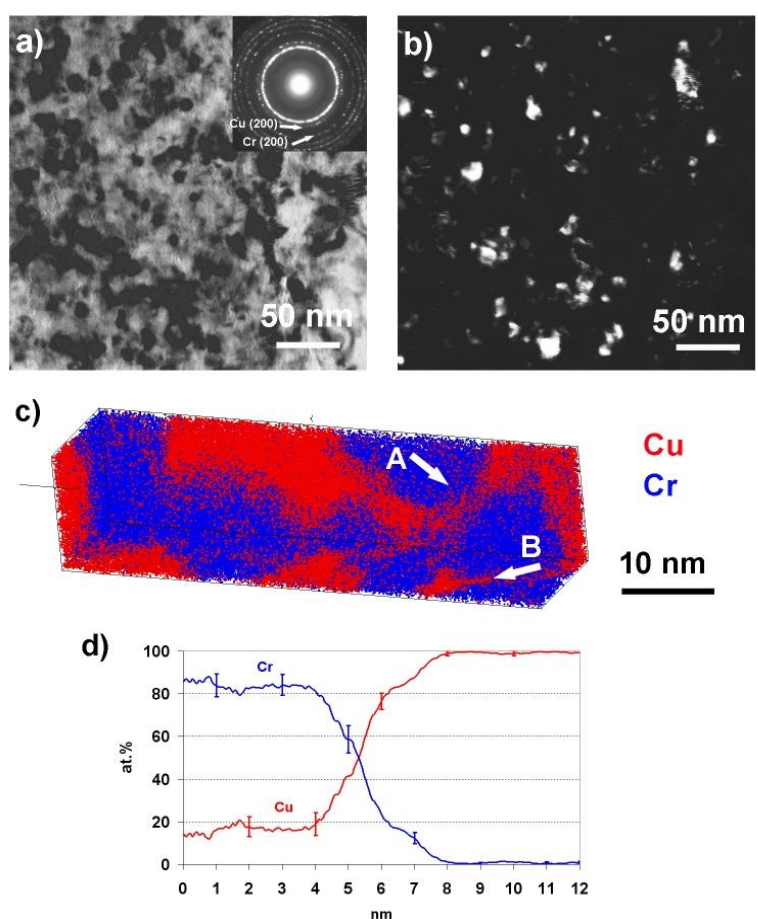

Figure 4
(a) TEM bright field image and related SAED pattern of a Cu57Cr43 composite processed by HPT up to 25 revolutions; (b) Dark field image of the same area obtained by selecting some (111)Cu and (110)Cr lattice reflections (first Debye-Scherrer ring); (c) 3D reconstruction of an analysed volume (12x12x50 nm3) showing the distribution of Cu and Cr. (d) Composition profile computed across a Cu/Cr interface showing that the interface is not chemically sharp and that some SPD induced diffusion of Cu occurred (sampling volume thickness 1 nm) – [84]

In the case of the CuCr composite processed by HPT, APT data clearly revealed some non-equilibrium interdiffusion (Fig. 4). Interfaces are not chemically sharp and some Cu is transported across the Cu/Cr interfaces forming Cu supersaturated solutions in the bcc Cr phase. Similar features were reported in Cu-Fe [108, 109], Cu-Co [110] and Al-Ni [88] binary systems. Such mechanical mixing does occur only in the nanoscale regime and is thus probably promoted by the high interfacial energy and SPD induced vacancies that may promote the atomic mobility [108]. However, some other authors do believe that dislocations play a critical role and are underlying the so-called "kinetic roughening" model [111]. Atoms would be shifted across interfaces by the shear of atomic glide planes. The final state would be determined by the balance between these forced jumps and decomposition due to thermal diffusion. Recent APT measurements on the Cu-Ag system seem to validate this approach [112].

In summary, SPD produces high densities of lattice defects and refines the size and distribution of second phases through fragmentation by intense shear. SPD can also introduce supersaturation of alloying elements or precipitation of fine particles during subsequent aging.

## 5. Diffusion along grain boundaries in ultrafine grained materials

As it was stated above (section 2) the diffusion investigations are a highly sensitive probe for investigation of structural modifications on the atomic scale since the thermally-activated diffusivity depends exponentially on the corresponding activation barriers which are determined by the interatomic potentials and the atomic environment. Thus, dedicated measurements of the atomic mobility at low temperatures, when diffusion within undisturbed regions of the crystal lattice is frozen, can be used for analyzing the structural modifications of short-circuit diffusion paths, such as grain boundaries.

The preceding analysis demonstrated that SPD processing modifies structure (see section 2 & 4) and thermodynamics of interfaces introducing various defects (e.g. abundant GB vacancies and dislocations at and near interfaces) and (optionally) inducing segregation (section 3). These structure modifications affect the kinetic properties of interfaces, which is the subject of the present analysis.

First direct measurements of grain boundary diffusion in severely deformed materials yielded ambiguous results – both similar [113] and enhanced [114-116] rates of atomic transport were deduced with respect to the grain boundary diffusivities in reference coarse-grained materials. Systematic measurements by the radiotracer technique discovered a hierarchic nature of internal interfaces which are developing as a result of strong dislocation activity during SPD processing and, presumably, of a localization of plastic flow [12, 117]. Both "conventionally fast" as well as "ultra-fast" short-circuit diffusion paths were observed in SPD processed materials, with the latter being embedded in a network of grain boundaries akin relaxed high angle GBs as they exist in annealed coarse-grained materials (where these boundaries constitute the fastest short-circuit diffusion paths) [12]. This is an important discovery of the radiotracer method which provides sample-averaged information. The existence of a hierarchy of interfaces in plastically-deformed metals has been pointed out by Hansen and co-workers [118, 119] by introducing the so-called extended or geometrically necessary boundaries (GNBs) and incidental dislocation boundaries (IDBs). However, the diffusion studies indicate another type of the hierarchy which corresponds to other kinds of involved interfaces, since the diffusivity of dislocation or low-angle dislocation grain boundaries is definitely lower than that of general high-angle interfaces [116, 120]. Generalizing these findings, the following hierarchy of interfaces in SPD materials can be proposed (in the order of decreasing diffusivities):

- non-equilibrium interfaces (probably of different types and representing a certain spectrum of diffusivities and structures);
- general high-angle grain boundaries (with diffusivities and, probably, structure being similar to those of relaxed high-angle grain boundaries);
- highly defected (non-equilibrium) twin boundaries with diffusivities similar to those of the previous level [121]. Note that diffusion along relaxed twin boundaries is hardly measurable;
- low-angle boundaries, dislocation walls, single dislocations.

A comprehensive theory of SPD processing and grain refinement has to include all these levels of the hierarchy, which on the other hand may critically depend on processing routes and regimes (temperature, strain rate, applied pressure and so on). It is important that these levels correspond to different scales with the mesh size ranging from several micrometers (the non-equilibrium boundaries) down to hundred (dislocation walls) or even tens of nanometers (nano-twins). The appearance of general high-angle grain boundaries (with properties similar to those of relaxed interfaces in the coarse grained counterparts) in SPD materials depends obviously on processing temperature because this fact might be a clear indication of dynamic recovery processes during SPD.

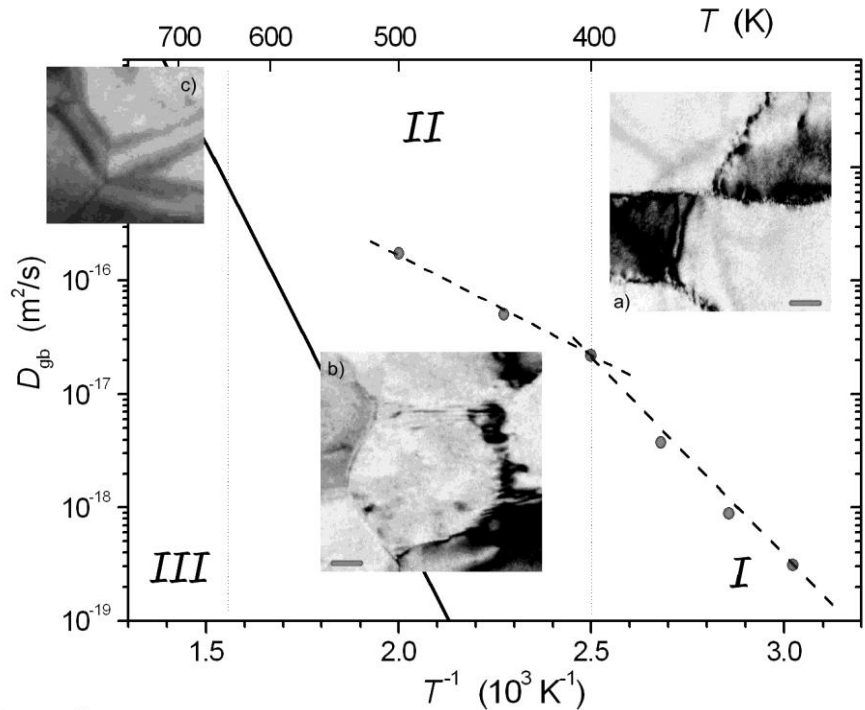

**Figure 5**
Correlation of typical grain boundary structures as observed by TEM (inserts a–c) and of diffusion enhancement in ECAP Ni (circles and dashed lines) [27]. The TEM images correspond to as-prepared state (a) and to typical structures observed after annealing treatments at 400 K (b) and 700 K (c) for 17 h. The scale bars correspond to 20 nm. The solid line represents the reference Ni grain boundary self-diffusion data in coarse-grained material [52]. There temperature intervals can be distinguished: I, II, and III, see text.

In Figure 5, the results of radiotracer measurements of grain boundary self-diffusion in UFG Ni (3N8 purity) after ECAP-processing (circles and dashed lines, [27]) are compared to the grain boundary diffusivity values that were obtained on a coarse-grained, polycrystalline material with a relaxed (annealed) grain boundary structure and a grain size of about 100 μm (the solid line, [52]). Clearly, short-circuit diffusion is significantly faster in UFG Ni than in coarse grained Ni and this diffusion enhancement depends critically on the temperature interval.

At lower temperatures, below about 400 K (region *I*), an almost linear Arrhenius behaviour is measured, with the corresponding activation enthalpy being roughly half of that which characterizes the coarse-grained Ni material. The experimental data at the temperatures above 400 K show a cross-over to a distinctly different yet consistent Arrhenius dependence (region *II*). This fact points to the attainment of a partially relaxed state of the non-equilibrium grain boundaries with a different metastable structural configuration as compared to that produced during SPD at room temperature and which is kinetically stable over a significant interval of annealing temperatures. The interval *III* corresponds to relaxation of the non-equilibrium state of GBs and overlapping processes of recrystallization / grain growth.

The regions *II* and *III* appear to be well separated in ECAP Ni due to the relatively low purity of the material used (99.6wt%). As a result of GB segregation of residual impurities the UFG microstructure turned out to be relatively stable and no significant recrystallization / grain growth was detected below 600 K [27].

The extremely fast tracer penetration in severely deformed pure Ni is consistent with the previous results of radiotracer diffusion measurements on ultra-fast transport in SPD-processed pure Cu [122-124] and Cu-based alloys [125-126][2]. On the other hand, such a clear kink in the Arrhenius dependence was not observed for grain boundary diffusion in UFG pure Cu and Cu-based alloy. This fact correlates with a lower homologous temperature of SPD processing (i.e. $T/T_m$, where $T_m$ is the melting point) for nickel ($T/T_m=0.17$) in comparison to that for copper ($T/T_m=0.22$), since the SPD process was carried out at room temperature in all cases under consideration. The low homologous temperature of deformation reduces the dynamic recovery processes which affect the concentrations of point defects, impurity segregation, atomic transport along interfaces and result in modified GB structures.

Such an observation substantiates the complexity of 'non-equilibrium GBs'. We have to admit that in addition to common parameters required to specify a relaxed high-angle grain boundary (i.e. the misorientation, inclination and the translation vector), extra parameters have to be introduced to characterize the non-equilibrium state. In a simplest approximation one may think of the defect density and/or the free volume density. Kinetic parameters, e.g. relaxation time(s), and thermodynamic parameters, e.g. segregation, compound formation or chemical ordering/disordering effects at interfaces, might also be involved in view of an inherent metastability of these interface states.

The region II substantiates a specific structure state of grain boundaries in SPD processed materials. It is important to note at this point, that a direct comparison of the diffusion measurements, which represent a macroscopically averaging method, with the highly local microstructure analyses results obtained by TEM-based techniques is not feasible. Yet, in all cases where SPD processed material was studied, similar contrast features as in Fig. 2(a) were observed by TEM and similar fast and ultrafast contributions to grain boundary diffusion were found. This correlation is sketched in Fig. 5 for ECAP Ni. The zipper contrast at the majority of GBs and strain contours around these GBs in the as-prepared state correlate with significant enhancement of interface diffusivity (region *I*). In the temperature interval *II*, the kink in the diffusion rate correlates with partial relaxation of the zipper contrast while the bulk strain/stress state is conserved to a large extent. Only in the interval *III* the recrystallization / grain growth processes trigger interface relaxation and recovery of the GB diffusivity. We conclude that the combined results on kinetics and structure of interfaces indicate that the non-equilibrium grain boundaries, revealing an increased width, a high density of localized defects (yet to be quantified) and high residual strain levels associated with them, possess a significantly enhanced diffusivity.

---

[2] An important difference is the absence of so-called percolated porosity in SPD Ni, which was discovered in ECAP Cu [127] and Cu-based alloys [128]. This porosity represents an extremely fast transport path toping the above introduced list of the short-circuit paths for diffusion in UFG materials.

Figure 5 indicates clearly that the term 'non-equilibrium grain boundary' encompasses a wide range of different states of interfaces with basically different kinetic/structure properties, cf. regions *I* and *II*. Remarkably enhanced diffusivities (although with significantly different effective activation enthalpies) and specific grain boundary structures observed in regions I and II substantiate a non-equilibrium state of interfaces in the corresponding temperature intervals. However, the pertinent interfaces reveal different TEM contrasts with presumably different strain/stress levels and defect populations. In the particular case of ECAP Ni, deformed at room temperature, we may talk about at least two distinct states of the non-equilibrium interfaces. The fundamental questions arise. What is common between the states *I* and *II*? Which properties have to be used for an unambiguous definition of the non-equilibrium state? The atomistic/structure reasons of the diffusivity enhancement also have to be understood.

The high diffusion rates are believed to be related to higher excess free energies of non-equilibrium grain boundaries in severely deformed Ni. Adopting the semi-empirical Borisov formalism [129] the excess free energy of non-equilibrium interfaces in ECAP Ni was found to be about 30% larger than in the annealed coarse-grained material [27], whereas about 10% increase was reported for SPD Cu [121].

We propose to generalize this formalism and to use the above-mentioned ansatz as *a measure of the excess free energy of interfaces irrespective of their state*. This phenomenological model gives rise to a *definition* of the non-equilibrium state of an interface with respect to relaxed general high-angle grain boundaries and furthermore introduces a convenient *measure* for the non-equilibrium state. In a series of papers by Nazarov et al. [15, 16] the non-equilibrium state of interfaces in SPD materials was related to the content of the extrinsic grain boundary dislocations (see section 2). It is interesting that both structure and kinetic approaches can be combined proving an extensive characterization of ECAP Ni [28]. The cross-over in the diffusion behaviour at 400 K, Fig. 5, correlates with a characteristic change of the relaxation time of the array of extrinsic grain boundary dislocations calculated according to Ref. [15] (and with the change of a typical HRTEM contrast at the interfaces) [27]. Whereas the agreement is encouraging for the region *I* in Fig. 5, the diffusion approach allows characterizing the state of interfaces in other regions where the dislocation approach of Nazarov et al. [15] fails since it predicts fully relaxed grain boundaries.

We have to admit that along with arrays of extrinsic grain boundary dislocations other defects should be included too. The following processes/phenomena contribute to the non-equilibrium state of grain boundaries in SPD processed materials:

- abundant vacancies and vacancy-like defects in interfaces produced by severe deformation;
- redistribution of the related excess free volume, release of local strains/stresses
- chemical effects (ordering) may be important in alloys and compounds affecting the atomic redistribution and retarding e.g. the stress/strain relaxation;
- segregation can be especially important in alloys involving even 2D compound formation along interfaces (see section 3).

The effective activation enthalpy of interface diffusion in ECAP Ni in the region *I* is similar to the effective activation enthalpy, which was found for recovery of vacancies in the material by DSC, suggesting that interface self-diffusion in as-prepared UFG Ni is governed by deformation-induced vacancies or vacancy-like defects. Basically, redistribution of these defects along with the stress relaxation in grains determined the transition from state *I* to state *II* with increasing temperature.

There is a striking correlation between the above mentioned kink in the Arrhenius dependence of Ni GB self-diffusion and the annihilation of single vacancies in SPD-processed Ni which was reported to occur at about 400 K, see e.g. [27, 130]. This fact indicates that the redistribution of vacancy-like defects – including their annihilation at GBs – results in a partial relaxation/transformation of the deformation-induced non-equilibrium state of the interfaces with lowering of the effective activation enthalpy of GB diffusion.

In order to give further insights into grain boundary structures affected by SPD processing, grain boundary diffusion of substitutionally (Ag) and interstitially (Co) diffusing solutes was investigated in coarse-grained (CG) as well as in UFG α-Ti produced by ECAP [131].

Co is known to be a so-called "ultrafast diffuser" in the crystalline bulk of CG α-Ti [132]. It occurred that Co is an ultrafast diffuser in grain boundaries of CG α-Ti, too. Astonishingly, at least at a first sight, grain boundary diffusion of Co in UFG α-Ti is slower than the interface diffusion in CG α-Ti while for Ag as diffusing species, grain boundary diffusion in the UFG material is significantly faster [131]. Due to SPD processing, a high concentration of defects, including those at grain boundaries, is created. The associated excess free volume offers effective (substitutional) traps for interstitially diffusing Co atoms. On the other hand, the Ag diffusivity is dramatically increased in UFG α-Ti as a result of severe plastic deformation. It is assumed that the physical origin of the increased diffusivity of Ag in UFG α-Ti is the formation of non-equilibrium GBs during the deformation process and the increase of the excess free volume of the interfaces.

In general the diffusion along non-equilibrium grain boundaries depends on the diffusion mechanism of the tracer. Interstitially diffusing atoms are trapped or scattered due to the high concentration of lattice defects in grain boundaries, which were induced by the severe plastic deformation. Accordingly, the interstitial diffusivity of such kind of elements can be slowed down.

A recent study by E. Schafler et al. [133] using X-ray line profile analysis showed that only half of the deformation-induced vacancies remain after unloading material from the high-pressure conditions maintained during the deformation. Thus, a significantly higher vacancy concentration is present in the material during deformation and this may potentially result in a much higher atomic mobility during the SPD treatment. Indeed, even if the very high applied pressure might hinder the migration of vacancies, the atomic mobility is the product of vacancy concentration and the vacancy mobility. Since the vacancy concentration is significantly increased during SPD, the atomic mobility could be also increased. The experimental observations of SPD induced segregations or enhanced particle dissolution are nice examples of enhanced atomic mobility (see section 3 and 4). Since grain boundary diffusion measurements were performed *ex-situ* (after the deformation and the high pressure has ceased), it is believed that the atomic mobility could be even higher during SPD. However, It remains unclear, if due to the increased value of the vacancy migration enthalpy under high hydrostatic pressure [134] the distribution of vacancies is rather homogeneous during the severe deformation.

Yet, due to the presence of a thermodynamic driving force, preferential diffusion of vacancies to the most potent sinks that are within diffusion distance, i.e. towards the high angle grain boundaries, can be safely assumed[3]. This scenario would lead to a composite structure with potentially more compliant GB regions and strong grain interiors.

## 6. Summary and Outlook

The results presented in this paper provide a strong evidence that SPD-processing synthesizes material with a significant fraction of high angle grain boundaries that possess higher excess free energy density, enhanced atomic mobility along the boundary plane, significant residual strain fields located at the near-boundary region and strongly increased segregation at the boundary and in the near-boundary region. These observations agree with early experiments [135, 136] and models [15] of non-equilibrium grain boundaries that suggested that SPD processing can trigger the interfaces to attain a non-equilibrium state as a result of heavy interaction with dislocations or imposed plasticity constraints.

However, the results also indicate that the final state of GBs created by SPD depends either on the dynamic recovery processes occurring in the vicinity of the boundaries, but also on possible interactions between lattice defects and impurities and/or solute elements. This could lead to a large variety of GBs, exhibiting various roughness, strain distribution, misorientation, local defect density, and these GB features can play a significant role in the properties of UFG materials. In addition, these features are closely related to the SPD processing regimes (temperature, strain rate and degree, applied pressure).

Due to their inherent large amount of GBs, the properties of UFG materials are rather sensitive to GB structures. For example, the thermal stability is one of the most important parameters for various applications (especially for creep and superplastic properties that have been demonstrated for various UFG alloys processed by SPD). It is interesting to note that on this point, GB structure and segregations might have a significantly favourable effect, as demonstrated for nanocrystalline Ni obtained by electrodeposition [38] and UFG Cu processed by ECAP [53]. Moreover, it seems that GB segregations could also affect the deformation mechanisms, leading in some cases to an increase of the strength that deviates from the Hall-Petch law [54]. Although the exact underlying mechanisms are not fully understood in a more general way, Molecular Dynamics (MD) simulations have shown that small composition fluctuations along grain boundaries may have a significant influence on the mechanical behaviour of bulk nanostructured materials [55]. In particular, the model suggested in [54] that is based on the influence of GB segregations on the formation of dislocations at grain boundaries enabled to evaluate the values of activation volume and strain-rate sensitivity of flow stress that agreed well with experimental data measured for the UFG Al alloy.

---

[3] yet, abundant vacancies and interstitials are created and annihilated during dislocation climb providing an important vehicle for deformation. The diffusion distances are small for such processes (about interatomic distances) and these defects can easily be removed. This effect may explain partially the observations of Schafler [133].

Moreover, so far the observation of the different properties of these high energy grain boundaries that have been measured either with local probes or stem from macroscopically averaging techniques, have yet to be shown to refer to one specific type of grain boundaries. The recent results obtained on the properties of grain boundaries in SPD-processed materials have also shown that the relaxation behaviour of these high-energy boundaries is far more complex than proposed by the earlier models of non-equilibrium grain boundaries.

It should be noted at this point, that there exists an apparent disagreement between the diffusional and structural (as measured by GPA) GB width (sections 2&5), on the one hand, and the chemical width determined by segregation studies (section 3), on the other hand. Whereas the structure and diffusional grain boundary widths represent well-known quantities, see e.g. [137, 52], the width of the segregated layer as a measure of non-equilibrium state, e.g. in a nominally single-phase material, has to be treated with caution. There is also another apparent contradiction between diffusion measurements giving the evidence of a fastest percolating pathway forming a mesh with a typical size corresponding to 5-10 grain diameters, while published bright-field TEM data taken at low resolution [135, 136] indicated a much larger proportion of "non-equilibrium" grain boundaries. It seems realistic to imagine that grain boundaries with a spectrum of excess free energy densities exist and that not all non-equilibrium grain boundaries affect diffusion in the same way such that a hierarchy among boundaries exists in SPD materials with respect to atomic mobility, residual stresses etc.. Some of these boundaries might yield specific contrast features in TEM (but one should note also that boundaries may also relax during thin foil preparation), some of them (or others) might reveal extra fast atomic mobility along the boundary plane by diffusion experiments, while there might exist some non-equilibrium state that has never been characterized yet by any technique. Presently, it seems still utopian to imaging an experiment in which diffusivity, atomic structure and segregation behaviour would be measured for the same non-equilibrium interface with reasonable accuracy, although such measurements are feasible in the case of bicrystals.

In this respect, it is of current concern to introduce quantitative parameters describing non-equilibrium grain boundaries. This will serve as the basis for creating the classification of different types of non-equilibrium GBs, which is important for further developing the concept of GB engineering of UFG materials towards the controlled enhancement of their properties. One such quantitative parameter could be defined on the basis of the relative excess free energy density with respect to a relaxed random high angle grain boundary. We have proposed a definition of the non-equilibrium state of interfaces in terms of a measurable quantity – the associated diffusion rate. There are a number of other parameters – the level of lattice distortion near an interface, the width measured by GPA, or the chemical width – which may be proposed to quantify the non-equilibrium state.

Moreover, the role of the non-equilibrium grain boundaries concerning the observed performance of SPD processed materials with ultrafine grained microstructure, especially concerning their mechanical properties, is still ubiquitous. Should the performance of the materials be described by a composite model consisting of grain interiors, grain boundaries and non-equilibrium grain boundaries? Further studies on model materials applying several different measurement methods on the very same specimens might allow addressing these important issues in the future.

This paper substantiates the importance of a combined research which should include measurements of a number of the above-mentioned properties in the same material including their relaxation behaviour. One of the most striking and unique feature of SPD-processed materials is the creation of a hierarchy of fast diffusion pathways with significantly different atomic mobilities. This result of detailed measurements of the grain boundary diffusion rate is likely to reflect the different types of grain boundaries after SPD processing that cover different types of non-equilibrium grain boundaries as well as relaxed high angle grain boundaries. So far, the creation of percolating networks of pathways with specific atomic mobilities as well as the interaction of this microstructural feature with the macroscopic properties and performance of the materials is yet to be analyzed.

The use of strongly segregating impurities in UFG materials with non-equilibrium interfaces may provide another attractive route for property tuning by desired segregation of the selected elements. Most deep traps with high segregation energies will be filled by (oversized) solute atoms *reducing* the grain boundary energy. One may consider a scenario that the excess free energy of structural non-equilibrium interfaces becomes zero due to segregation and these defects will be thermodynamically stable. The kinetic properties of such interfaces would represent a highly topical subject as well as the response of the modified UFG material on mechanical as well as functional properties.

In this paper, the analysis was mainly focused on the non-equilibrium state of general high-angle GBs in SPD materials. Following [138], special GBs can be introduced in UFG materials to tune the desired properties. We note that twin boundaries can also be triggered to a non-equilibrium state by severe deformation [139] that might be crucial for the combination of high strength and reasonable ductility [140].

The concept of GB engineering or GB design was introduced by T.Watanabe in [138] where it had been proposed that the properties of polycrystalline materials may be effectively changed by deliberate and careful tailoring of grain boundary character distribution. The concept of grain boundary engineering can be developed and applied to SPD materials since a wide variety of GBs could be achieved by variation of SPD processing routes and regimes [141, 142]. Indeed, low and high angle grain boundaries with various proportions could be obtained, including some segregations of solute elements or nanoscaled precipitates that could dramatically change the properties like the thermal stability. Therefore, the design of specific grain boundaries for optimal properties could be considered in the near future. Such an approach may open a new area for bulk nanostructured materials.

Thus, one can expect that new advanced properties may be achieved through variations of GB structure in the ultrafine-grained materials processed by SPD and these studies, in the authors' opinion, should become a dynamic trend of further research in the field of nanomaterials and nanotechnologies.

**Acknowledgements**

The authors gratefully acknowledge funding by CNRS, DFG and RFBR of a French-German-Russian Tri-Lateral Seminar on "Atomic Transport Kinetics in Bulk Nanostructured Materials", Rouen (France), May 2010, which nucleated the present overview paper. The authors also acknowledge individual funding on the above research topic by CNRS, DFG and RFBR.